\documentclass[aps,prb,twocolumn,superscriptaddress]{revtex4-1}

\usepackage{amssymb}
\usepackage{graphicx}
\usepackage{amsmath}

\usepackage{hyperref}

\begin{document}
	
	\title{Dispersive SYK model: band structure and quantum chaos}
	\author{Pengfei Zhang}
	\affiliation{Institute for Advanced Study, Tsinghua University, Beijing, 100084, China}
	\date{\today}
	
	\begin{abstract}
	The Sachdev-Ye-Kitaev (SYK) model is a concrete model for non-Fermi Liquid with maximally chaotic behavior in $0+1$-$d$. In order to gain some insights into real materials in higher dimensions where fermions could hop between different sites, here we consider coupling a SYK lattice by a constant hopping. We call this dispersive SYK model. Focusing on $1+1$-$d$ homogeneous hopping, by either tuning temperature or the relative strength of random interaction (hopping) and constant hopping, we find a crossover between a dispersive metal to an incoherent metal, where dynamic exponent $z$ changes from $1$ to $\infty$. We study the crossover by calculating  spectral function, charge density correlator and the Lyapunov exponent. We further find the Lyapunov exponent becomes larger when the chemical potential is tuned to approach a Van Hove singularity because of the large density of states near the Fermi suface. The effect of the topological non-trivial bands is also discussed.
	\end{abstract}
	
	\maketitle
	
	\section{Introduction}
	
Sachdev-Ye-Kitaev model, proposed by Kitaev \cite{Kitaev2} and closely related to the work by Sachdev and Ye \cite{SY}, describes $N$ modes Majorana fermions $\chi_i$ with random interaction on a quantum dot. The model is studied extensively on both field theory side  \cite{Kitaev2,Comments,spectrum1,spectrum2,spectrum3,Liouville,Liouville2,SYK new,SYK new2,SYK new3,SYK new4,quench1} and gravity side \cite{Comments,bulk Yang,bulk spectrum Polchinski,bulk2,bulk3,bulk4,bulk5,syk-bh,SYK g new1,SYK g new2,SYK g new3,new g,new g2}. In this paper, we will focus on the the field theory description and the Lagrangian in imaginary time is given by:
\begin{align}
L_{SYK}=\frac{1}{2}\sum_i\chi_i\partial_\tau\chi_i+\frac{1}{4!}\sum_{ijkl}^{N}J_{ijkl}\chi_i \chi_j \chi_k \chi_l. \label{SYK}
\end{align}
After canonical quantization, the commutation relation is given by $\{\chi_i, \chi_j \}=\delta_{ij}$. Here $J_{ijkl}$ are some independent random Gaussian variables with distribution:
\begin{align}
\overline{J_{ijkl}}=0\ \ \ \ \overline{J_{ijkl}^{2}}=\frac{3!J^2}{N^3}.
\end{align}

This model can be solved in the Large-$N$ limit. Moreover, if we focus on the low energy limit $\beta J\gg1$ where the interaction term dominates, there is an emergent conformal symmetry under transformation $\chi_i(\tau)\rightarrow\left(\frac{d\tau '}{d\tau}\right)^{1/4}\chi_i'(\tau')$. This symmetry largely simplifies analytical calculations and makes the model an ideal platform to check new ideas on non-Fermi Liquid, quantum chaos and holographic duality:

Firstly, by solving the Schwinger-Dyson equation for Majorana operator, SYK model is shown to be a concrete model for non-Fermi Liquid \cite{Comments,Kitaev2}:
\begin{align}
\langle\mathcal{T}_{\tau}\chi_i (\tau)\chi_j (0) \rangle=G(\tau)\delta_{ij}=b\frac{\textrm{sgn}(\tau)}{|\tau|^{\frac{1}{2}}}\delta_{ij}.
\end{align}
After a Fourier transformation, one finds the spectral function diverges as $\omega^{-1/2}$ as $\omega\rightarrow0$, which is a signature of non-Fermi Liquid. 

Secondly, if one use the out-of-time-ordered correlation function to define a quantum Lyapunov exponent $\lambda_L$:
\begin{align}
\langle \chi_i(t-i3\beta/4)\chi_j(-i\beta/2)&\chi_i(t-i\beta/4)\chi_j(0)\rangle_\beta\notag\\&\sim-\exp(\lambda_L t)\ \ \ (t\gg \beta).
\end{align} 
The model is shown to be maximally chaotic with $\lambda_L=2\pi/\beta$ \cite{Comments,Kitaev2}, which is proved to be a bound for a generic quantum system \cite{bh1,bh2,bh3,prove}. Conjectured by Kitaev \cite{Kitaev2}, this maximally chaotic behavior also implies a holographic bulk description. Indeed, the effective action for the SYK model, which is largely determined by the symmetry of the system, is the same as some dilaton gravity in nearest AdS2 space-time \cite{bulk Yang}.

Following the basic construction of SYK model, many different generalizations is proposed to extend our knowledge of non-Fermi Liquid, quantum chaos and holographic duality \cite{numerics wenbo,Yingfei1,Yingfei2,generalization 1,wenbo susy,susy2,no disorder1,no disorder2,no disorder3,no disorder4,no disorder5,no disorder6,no disorder7,no disorder8,no disorder9,no disorder10,Altman,thermal transport,high-D1,high-D2-con,yyz condensation,sk jian,generalization 2,transition1,our,Balent,thick,no disorder new,susy3,our2,new gen1,new gen2,new gen3,new gen4,new gen5,new gen6,new gen7,new gen8,new gen9,new gen10,new gen11,new gen12,new gen13,new gen14}. One important strategy is to couple different quantum dots (with SYK interaction or not) to get a model in higher dimension or study the transition (or crossover) between different fixed points \cite{Altman,Balent,our,our2,Yingfei1,Yingfei2,thermal transport,sk jian}. In \cite{Balent}, the authors study a lattice of complex fermion version of SYK models with intra cell random interaction and inter cell random hopping:
\begin{align}
H'=(\sum_{i,j,x}V_{ij,x}c^{\dagger}_{i,x}c_{j,x+1}+\text{h.c.})+\frac{J_{ijkl,x}}{4}c^{\dagger}_{i,x}c^{\dagger}_{j,x}c_{k,x}c_{l,x}.\label{H'}
\end{align}
Here $V_{ij,x}$ and $J_{ijkl,x}$ are some random numbers with expectation:
\begin{align}
\overline{J_{ijkl,x}}=0,\,\,\,\,\,\,\overline{V_{ij,x}}=0,
\end{align}
and their variances are 
\begin{align}
\overline{|J_{ijkl,x}|^{2}}=\frac{2J^2}{N^3},\,\,\,\,\,\, \overline{|V_{ij,x}|^{2}}=\frac{V^2}{2N}.
\end{align}

By a naive power counting, the low-energy physics is dominated by the random hopping (in \cite{Balent} they call this a heavy Fermi-Liquid) and for small $V/J$ there should be a crossover to the fixed point of weakly coupled SYK models (they call this an incoherent metal) when we increase the temperature. They study this crossover by calculating spectral function, entropy and transport coefficients including charge diffusive constant. Interestingly, they find the transport coefficients show similar behavior as a high-$T_c$ superconductor in its strange metal phase. An equivalent crossover is also studied by calculating spectral function, entropy and $\lambda_L$ in \cite{our}.

Surprisingly, in this random hopping model, although there is a coupling between different sites, the two-point correlation function $G(x,t)$ is non-zero only for $x=0$. As a result after a Fourier transformation, there is no dispersion for fermions. This is a consequence of the disorder hopping which only couples the fermion operators at the same site after disorder average. On the other hand in real materials, dispersion, or band structure, indeed plays an important role. For example, the nesting effect leads to instabilities such as charge-density wave and BCS superconductivity, which is the origin of numbers of different ordered phases \cite{shankar}. In $1$-D, it is also the nesting effect that lead to the celebrated Luttinger Liquid theory which describes another type of non-Fermi Liquid \cite{Q1D}. Thus the interplay between band structure and the chaotic random interaction is worth studying while up to now no model exists in the context of generalized SYK models.

In this paper we study the effect of dispersion on coupled SYK lattice by adding random interaction and random hopping between different copies of free fermions with constant hopping $t$ on a lattice. We call this dispersive SYK model. To be concrete, we focus on the $1+1$-d case with homogeneous hopping while the construction applies for any lattice. In the section 2, we study the self-consistent equation for two point correlation function and discuss results for spectral function, which clearly shows a crossover between dispersive metal to a super-local incoherent metal. Then, based on the two point correlation function, we study the correlation function for charge density in section 3. In the dispersive limit $t\gg V$ and $t\gg J$, the spectral for charge density shows two peaks near $\omega\sim c p$ because of the presence of density wave for metal in 1-D where $c$ is the velocity for this wave mode. These peaks crossover to a diffusive peak for incoherent metal near $\omega\sim0$. The behavior of the Lyapunov exponent when tuning $\beta$ for small $t/J$ is studied in section 4 and we further find if we tune the chemical potential to approach the Van Hove singularity, the Lyapunov exponent becomes larger. We also briefly discuss some effects of topological bands. At last we discuss some possible extensions of this work to be studied in the future.

\section{Dispersive SYK model and Green's function}

\begin{figure}[t]
	\centering
	\includegraphics[width=0.4\textwidth]{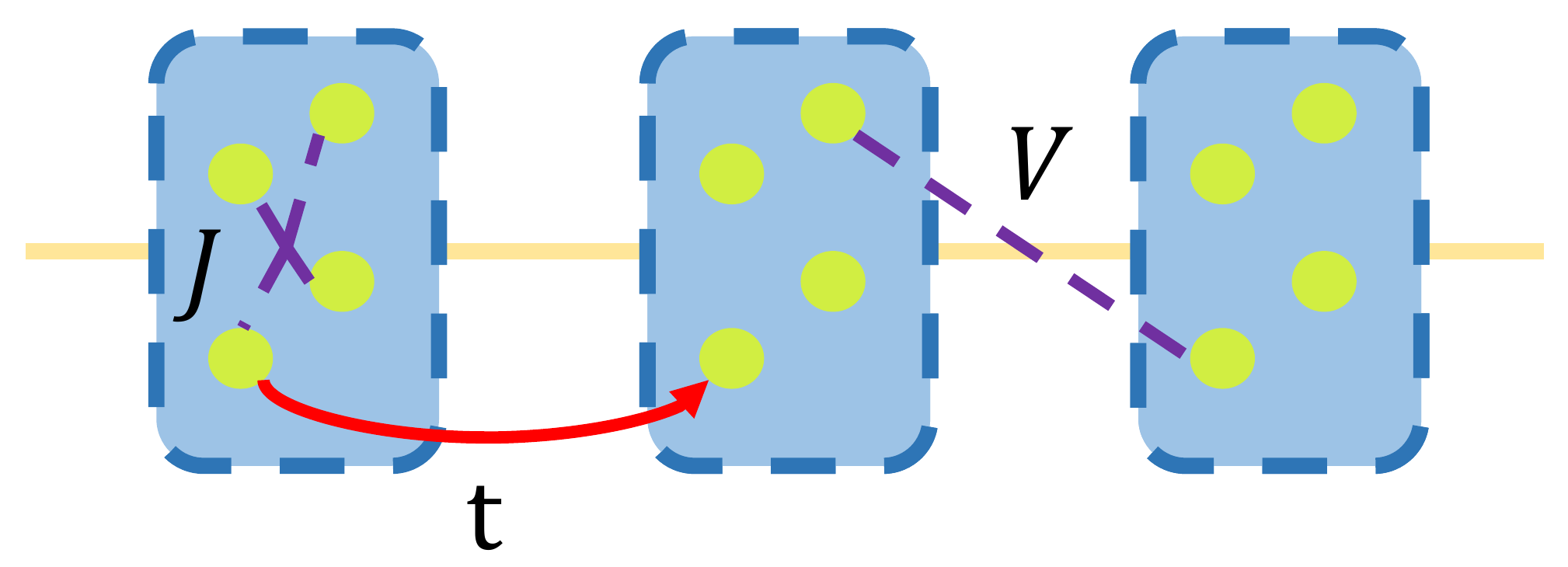}
	\caption{A pictorial representation of the model described by \eqref{H}. Different dots represent different modes in each unit cell. The dashed lines represent random interaction (hopping) and the solid line represent the constant hopping.}\label{9}
\end{figure}

As shown in Figure \ref{9}, We consider adding a constant hopping to a quadratically coupled SYK lattice \eqref{H'}:
\begin{align}
H=&(\sum_{i,j,x}V_{ij,x}c^{\dagger}_{i,x}c_{j,x+1}+\text{h.c.})-t(\sum_{i,x}c^{\dagger}_{i,x}c_{i,x+1}+\text{h.c.})\notag\\&+\sum_{i,j,k,l,x}\frac{J_{ijkl,x}}{4}c^{\dagger}_{i,x}c^{\dagger}_{j,x}c_{k,x}c_{l,x}-\mu\sum_{i,x}c^{\dagger}_{i,x}c_{i,x}.\label{H}
\end{align}
The distributions of $J_{ijkl,x}$ and $V_{ij,x}$ are the same as those in \eqref{H'}. To be concrete, we first restrict ourselves to the one-dimensional translational invariant chain with homogeneous hopping and then discuss how to generalize the result. $\mu=0$ corresponds to half-filling because of the particle-hole symmetry $c_{i,x}\rightarrow c^\dagger_{i,x}(-1)^i$ after disorder average. Here the summation over $i$ is from $1$ to $N$ and $x$ is an integer that labels the lattice sites. Because the constant hopping will make the fermions dispersive, we call this dispersive SYK model.

The constant hopping $t$ here only couples different sites with the same mode index $i$. This means \eqref{H} can be considered as some coupled wire model. The scaling dimension for $c_{i,x}$ at the fixed point of the $\text{SYK}_4$ is $1/4$ and then all quadratic terms are relevant. One thus expect at low temperature, the low-energy part of the theory is governed by random or constant hopping. For example, if one set $V=0$, for small $t/J$ at extreme low temperature, the system is in a dispersive metal phase with dynamical exponent $z=1$ while at higher temperature, it becomes a super-local non-Fermi Liquid with dynamical exponent $z=\infty$. One could also study the crossover with fixed temperature and tune $t/J$. In this paper, we will analyze the crossover by calculating spectral function, charge density correlator and the Lyapunov exponent.

We first study the problem using the imaginary-time path integral. Usually, one should use the replica trick to handle a problem with disorder. But because of the large-N suppression, the coupling between different replicas can be neglected to the leading order and thus we assume there is no spontaneous breaking of the replica permutation. Indeed this is the same argument as what we have in the original SYK model \cite{Comments}. From now on, we will simply drop the replica indexes.

We define $$G_{ij}(x,\tau;y,\tau')\equiv\overline{\left<\mathcal{T}_\tau c_{i,x}(\tau)c_{j,y}(\tau')\right>}.$$ After disorder average, there is no coupling between different modes $i$. Then this fact, together with the translational invariance, says $G_{ij}(x,\tau;y,\tau')=\delta_{ij}G(x,\tau;y,\tau')=\delta_{ij}G(x-y,\tau-\tau')$. The Schwinger-Dyson equation for $G$ is given by:
\begin{align}
G^{-1}(p,\omega_n)=-i\omega_n-\mu+\epsilon(p)-\Pi(p,\omega_n).\label{SDe}
\end{align}
Where $\epsilon(p)=-2t\cos p$ is the single particle dispersion and $\Pi$ is the self-energy by summing up all one-particle irreducible diagrams which can expanded in terms of $1/N$. To the leading order, it is given by diagrams shown in Figure \ref{melon}. Because of the disorder average, these diagrams only couple fermion operators at the same site:

\begin{figure}[t]
	\centering
	\includegraphics[width=0.4\textwidth]{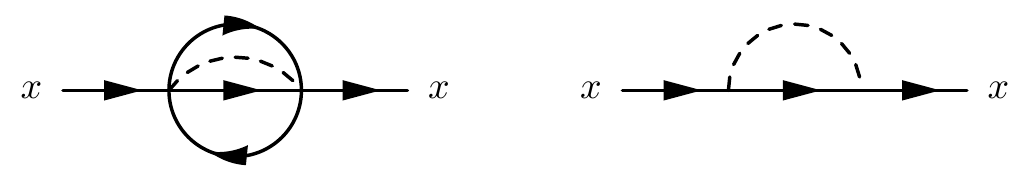}
	\caption{Self-energy in large $N$ expansion for two point correlation function. The dashed line indicates the disorder average and the solid line is the full-propagator for fermion.}
	\label{melon}
\end{figure}

\begin{align}
\Pi(x,\tau)=\delta_{x,0}(V^2G(\tau)+J^2G(\tau)^3).\label{SE}
\end{align}
Here $G(\tau)=G(x=0,\tau)$ where we have omitted $x=0$. From now on we take similar convention for any Green's function or spectral function. The locality of $\Sigma$ in $x$ means the only momentum dependence of $G(p,\omega_n)$ comes from $\epsilon(p)$. This set of equations is complete and could be generalized to any lattice with $\epsilon(p)$ replaced by the single-particle dispersion on it. One could solve the \eqref{SDe} and \eqref{SE} by numerics. But to obtain the spectral function, it is better to do the analytically continuation first which gives the self consistent equation for retarded Green's function directly:
\begin{align}
G_R(x,t)&\equiv-i\theta(t)\overline{\left<\{ c_{i,x}(t),c_{i,0}(0)\}\right>},\\
G^{-1}_R(p,\omega)&=\omega+\mu-\epsilon(p)-V^2G_R(\omega)-J^2\Sigma_R(\omega).\label{SDeGR}
\end{align}
Here $\Sigma_R(\omega)$ is the contribution from $\text{SYK}_4$ interaction by analytical continuation of $\int d\tau e^{i \omega_n \tau} G(\tau)^3$ after integration over $\tau$ \cite{Comments,Altman}:
\begin{align}
\Sigma_R(\omega)=-iJ^2\int_0^\infty dt e^{i\omega t}(n_1^2(t)n_2(t)+n_3^2(t)n_4(t)).\label{num1}
\end{align} 
with 
\begin{align}
&n_1(t)=\int A(\omega)n_F(-\omega)e^{-i\omega t}=n_4(t)^*,\\
&n_2(t)=\int A(\omega)n_F(\omega)e^{i\omega t}=n_3(t)^*.\\
&A(\omega)=-\frac{1}{\pi}\text{Im}G_R(\omega).
\end{align}

\begin{figure}[t]
	\centering
	\includegraphics[width=0.45\textwidth]{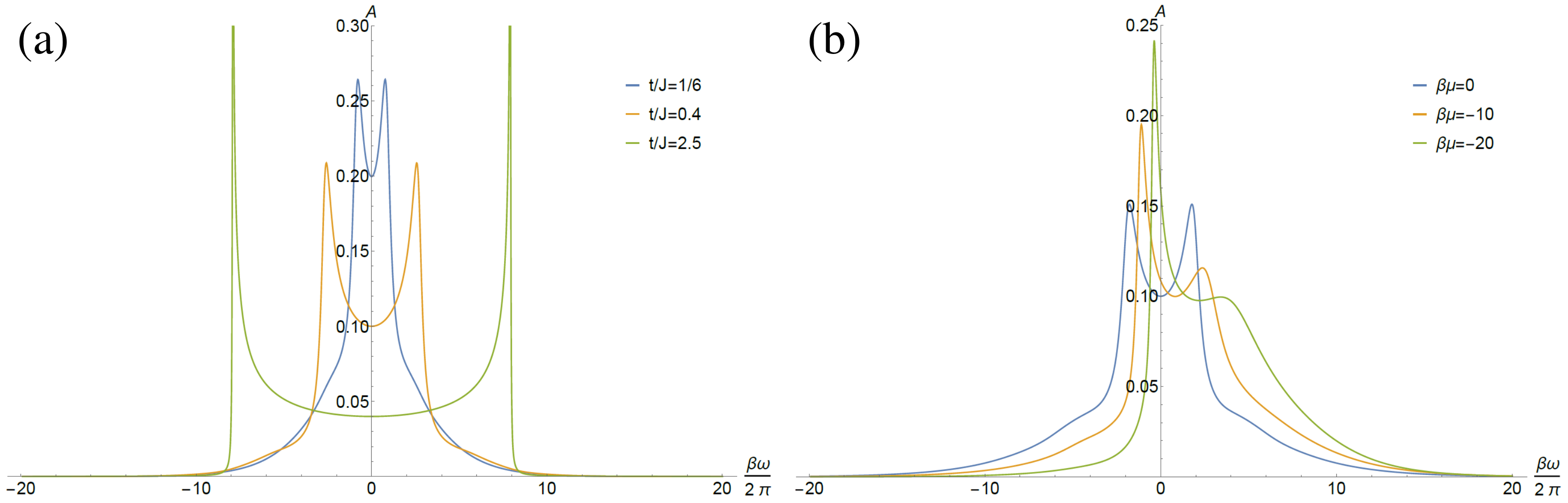}
	\caption{(a)Spectral function $A(\omega)$ by numerically solving \eqref{num1} and \eqref{num2} for different $\beta t/J$ with $V=\mu=0$ and $\beta t+\beta J=35$. There is an apparent particle-hole symmetry. (b)Spectral function $A(\omega)$ for different $\mu$ with $\beta t=10$ and $\beta J=50$.}\label{spectral}
\end{figure}

For one-dimensional chain, one could further simplify the problem by using the analytical result for the integral over $p$:
\begin{align}
G_R(\omega)=&\int \frac{dp}{2\pi}\frac{1}{a+2t \cos p}=\frac{1}{a-2t}\sqrt{\frac{a-2t}{a+2t}},\\
\text{with}&\ \ a=\omega+\mu-V^2G_R(\omega)-J^2\Sigma_R(\omega).\label{num2}
\end{align}
this equation then only contains the correlation function with no spatial distance and the square root is defined with a cut line along the negative real axis. After solving the solution, the correlation function for any momentum is then given by \eqref{SDeGR}. In fact we know that:
\begin{align}
G_R&(x,\omega)=\int \frac{dp}{2\pi}\frac{\cos(xp)}{a+2t \cos p}\notag\\&=\frac{1}{a-2t}\sqrt{\frac{a-2t}{a+2t}}\left(\frac{1}{2t}(-a+\sqrt{2t+a}\sqrt{-2t+a})\right)^{|x|}. \label{GRX}
\end{align}

Now we discuss the behavior of $A$ in different limits. For simplicity, we set $V=0$ and $\mu=0$ here. At finite temperature when $t\gg J$, we are studying a free hopping fermions and the spectral function $ A(\omega)$ is given by:
\begin{align}
A(\omega)=\int\frac{dp}{2\pi}\delta(\omega-2t\cos p)=\frac{1}{2\pi t|\sin p_*|}.
\end{align} 
for $-2t<\omega<2t$ with $p_*=\arccos(\frac{\omega}{2t})$. For $\omega\rightarrow\pm2t$ the density of states diverges. This is the location of the Van Hove singularity. For small but finite $J/t$, the singularity is broadened by interaction and becomes a peak with finite density of states. Then, if we consider larger interaction $J$, the spectral function is further broadened and at last it will approach the result of a SYK$_4$ model, which is a narrow peak near $\omega\sim0$ \cite{Comments}: 
\begin{align}
A(\omega)=A_{\text{SYK}}(\omega)\propto\text{Im}\left(-i\frac{\Gamma(1/4-\frac{i\beta\omega}{2\pi})}{\Gamma(3/4-\frac{i\beta\omega}{2\pi})}\right).
\end{align}

Numerical results for $A(\omega)$ are shown in Figure \ref{spectral}, which reproduces the result in both limit correctly. For simplicity, we set $\mu=V=0$. This numerical results also confirms the hopping is relevant at low-energy limit since the peak at small $\omega$ is always split. This means the constant hopping suppresses the low-energy density of states. We know that the Lyapunov exponent describes relative long time behavior and is thus governed by low-energy modes. We expect the system to be less chaotic after adding the hopping term. 

For free fermions, one method to increase the density of state near the Fermi surface is to tune the chemical potential. Thus we further study the spectral function with different chemical potential for this dispersive SYK model. When tuning $\mu\sim \pm2t$, we find the low-energy density of states becomes large, which is similar to a non-interacting model, and the Lyapunov exponent is expected to become larger. We will check this is indeed true in later sections.

\section{Crossover from dispersive metal to diffusive metal: charge density correlator}
Since SYK model is a concrete model for holographic duality \cite{bulk Yang}, the crossover from the dispersive metal to the incoherent metal may have a bulk version. It should be a crossover between different gravity theory with different dynamical exponent $z$ which is interesting on its own right. To make progress in finding the effective holographic dual of some theory, it is important to study some correlation function with universal behaviors. Here we choose to study the charge density correlator in field theory side. First we would like to recall the knowing result in both limit and then use the Keldysh approach to study the correlator in all regime.

In $1$-D and low-energy limit, free fermions can alternatively be described in term of density waves which is a well-defined quasi-particle with linear dispersion. This is called bosonization. From spectral function, (for a tight-binding chain with hopping strength $t$ and for simplicity we only consider the result for $\mu=0$ in this section, where the charge transport decouples from the energy transport. \cite{thermal transport,Balent}) it says that the retarded Green's function for charge density correlator $\Pi_{R,nn}(p,\omega)=-i\int dtdx\ e^{i\omega t-ik x}\theta(t)\left<\left[n(x,t),n(0,0)\right]\right>$ contains two poles for $t\gg\omega,p,T$:
\begin{align}
\Pi_{R,nn}^{DM}(p,\omega)=\frac{1}{2\pi}(-\frac{p}{\omega-2 t p+i\epsilon}+\frac{p}{\omega+2 t p+i\epsilon}).\label{LL}
\end{align}
The velocity for the density wave is given by $2t$ and $\epsilon$ is infinitely small for free fermions. As a result for large $t/J$ and $t/V$, one expect similar behavior for the dispersive SYK model with finite width $\epsilon$ because the random interaction leads to a finite lifetime for quasi-particles.

On the other hand, in the limit $t/J\rightarrow0$ and $t/V\rightarrow0$, the model is diffusive and is called an incoherent metal \cite{Balent}. $\Pi_{R,nn}$ is calculated by taking the phase fluctuation of fermions into account which gives the leading contribution in $1/N$ expansion. It is found that there is no well-defined quasi-particle for density operator and the behavior for $\Pi_{R,nn}$ is hydrodynamical:
\begin{align}
\Pi_{R,nn}^{HD}(p,\omega)=\frac{-KD p^2}{i\omega-D p^2}\label{HD}
\end{align}
where $K$ is the compressibility $\partial n/\partial\mu$ and $D$ is the diffusive constant for charge density. This corresponds to a single diffusive peak for the dynamical structure factor $S_{nn}(p,\omega)$ at $\omega\sim0$.

For intermediate value of $t/J$ and $t/V$, there should be a crossover between \eqref{LL} and \eqref{HD}. Interestingly, similar crossover also appears in a superfluid when tuning temperature where the second sound connects to a diffusive mode for entropy. We explore the crossover by using the Keldysh contour. There are two time contours $+/-$ and thus two copies of fields $c_{x,+}$ and $c_{x,-}$ (we drop $N$ modes index for simplicity because different $i$ decouples to the leading order of $N$). The phase fluctuation is introduced by $c_{x,\pm}\rightarrow \exp(-i\phi_{\pm}(x))c_{x,\pm}$ with an assumption that the dependence of $x$ for $\phi_\pm(x)$ is smooth. We choose the following convention \cite{Kamenev}: 
\begin{align}
&c_1=\frac{1}{\sqrt{2}}(c_++c_-),\ \ \ \ c_2=\frac{1}{\sqrt{2}}(c_+-c_-),\\
&\overline{c}_1=\frac{1}{\sqrt{2}}(\overline{c}_+-\overline{c}_-),\ \ \ \ \overline{c}_2=\frac{1}{\sqrt{2}}(\overline{c}_++\overline{c}_-),\\
&\phi_{\text{cl}/q}=\frac{1}{2}(\phi_{+}\pm\phi_{-}),\ \ J_{\text{cl}/q}=\frac{1}{2}(J_{+}\pm J_{-}),\\&N_{\text{cl}/q}=\frac{1}{2}(N_{+}\pm N_{-}).
\end{align}
Here $J_\pm$ are source terms added to the action by $$S[J_\pm(x)]=\int dt dx \left(J_+(x)N_+(x)-J_-(x)N_-(x)\right)$$ to extract the correlation function and $N=c^\dagger c$ is the charge density operator. One have the standard Green's function in Keldysh formalism: $$\left<c_{x,\alpha}(t)\overline{c}_{0,\beta}(0)\right>=\begin{pmatrix}
G_R(x,t)&G_K(x,t)\\
0&G_A(x,t)
\end{pmatrix}.$$
with $G_K(p,\omega)=(1-2n_F(\omega))(G_R(p,\omega)-G_A(p,\omega))$ for thermal equilibrium which is the fluctuation-dissipation theorem. The source term is given by:
\begin{align}
J_+N_+-J_-N_-=2J_{\text{cl}}N_q+2J_{q}N_{\text{cl}}.
\end{align}

A physical variation of chemical potential $\mu$ correspond to a symmetric change of $J$ on both contour: $J_+=J_-$ and similarly the physical density in given by $N_{\text{cl}}$. The retarded Green's function is then given by:
\begin{align}
N_{\text{cl}}=-\frac{i}{2}\frac{\partial \ln Z}{\partial J_q},\ \ \ \ 
\Pi_{R,nn}=\frac{\partial N_{\text{cl}}}{\partial J_{\text{cl}}}=-\frac{i}{2}\frac{\partial^2 \ln Z}{\partial J_{\text{cl}} \partial J_q}.
\end{align} 

In action $S$, after integrate out fermions, there are different terms that contribute to the  effective action of $\phi$ and $J$. They comes from the $\partial_t$ term, the constant hopping $t$ term and the random hopping $V$ term. The random hopping term, after introducing the phase $\phi_\pm(x)$ and only keep to the second order, it is given by:
\begin{align}
\mathcal{L}
=&\left(-V_{ij}c^\dagger_{x,+}(1-i\partial_x\phi_{+}-\frac{1}{2}(\partial_x\phi_{+})^2)c_{x+1,+}+\text{h.c.}\right)\notag\\&-\left(-V_{ij}c^\dagger_{x,-}(1-i\partial_x\phi_{-}-\frac{1}{2}(\partial_x\phi_{-})^2)c_{x+1,-}+\text{h.c.}\right).
\end{align}

Here we just focus on the $\phi_{\text{cl}}\phi_q$ part which contributes to the retarded correlator and keep up to the $\phi^2$ term. The contribution to effective action from this random hopping after taking the virtual process of particle-hole excitation into account comes from the two diagrams shown in Figure \ref{V term}: 

\begin{figure}[t]
	\centering
	\includegraphics[width=0.4\textwidth]{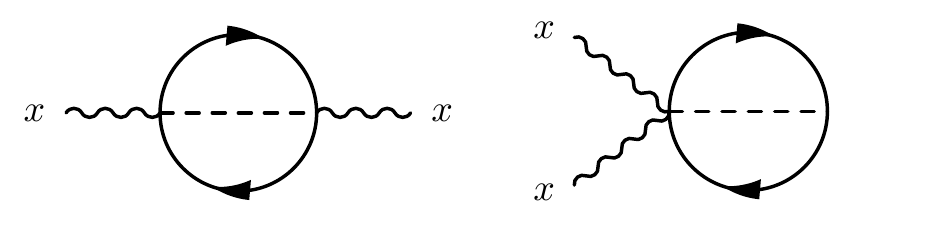}
	\caption{Contribution from random hopping $V$ for the phase fluctuation.}
	\label{V term}
\end{figure}

\begin{align}
iS^{(1)}_{V}&=\int d\omega dp \ D_r(\omega)\partial_x\phi_{\text{cl}}(p,\omega)\partial_x\phi_{q}(-p,-\omega).\\
D_r(\omega)&=-V^2\int\frac{dq_0}{2\pi} (G_R(q_0+\frac{\omega}{2})G_K(q_0-\frac{\omega}{2})\notag\\ &+G_K(q_0+\frac{\omega}{2})G_A(q_0-\frac{\omega}{2})-G_R(q_0)G_K(q_0)\notag\\
&-G_K(q_0)G_A(q_0)).
\end{align}

\begin{figure}[t]
	\centering
	\includegraphics[width=0.4\textwidth]{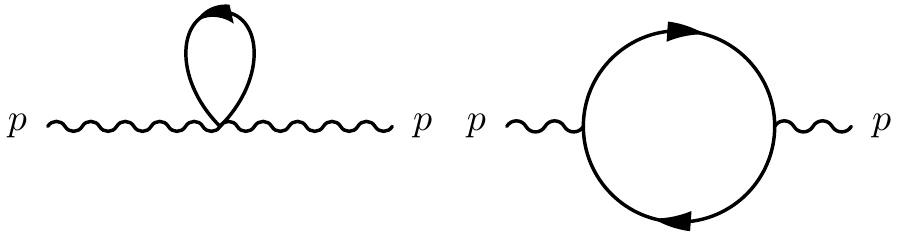}
	\caption{Contribution from the diamagnetic term and the current-current correlation.}
	\label{trans}
\end{figure}

This expression is consistent with the result by expanding the effective action in \cite{Balent} and $D_r$ is proportional to $\omega$ in low-energy limit. There are also contributions from $c\partial_tc$ and constant hopping $t$. Since we are interested in low-energy behavior, we approximate the current operator by the low-energy version:
\begin{align}
S_+=&(\partial_x\phi_+)(p)
\overline{c}_+(q+\frac{p}{2})c_+(q-\frac{p}{2})\frac{d\epsilon(q)}{dq}\notag\\
&-\frac{1}{2}(\partial_x\phi_+)(-p)(\partial_x\phi_+)(p)
\overline{c}_+(q)c_+(q)\frac{d^2\epsilon(q)}{dq^2}\notag\\&+\overline{c}_{x,+}(\partial_t\phi_++J_+)c_{x,+}.\label{trans S}
\end{align}
where we dropped some integrations over space or momentum for simplicity and keep only specific diamagnetic term that is relevant to the calculation later. A similar expression $S_-$ can be written for $-$ contour. For $t=0$, we only have the first term and the lack of momentum dependence also leads to some subtlety when one proceed and then one has to go back to imaginary path-integral \cite{Balent}. For finite $t$, one could do all calculation directly in the Keldysh contour. The second line of \eqref{trans S} contribute a diamagnetic term (Figure \ref{trans}):
\begin{align}
iS^{(2)}_{d}&=\int d\omega dp \ D_d\partial_x\phi_{\text{cl}}(p,\omega)\partial_x\phi_{q}(-p,-\omega).\notag\\
D_d&=-\int\frac{dq_0 dq}{(2\pi)^2}\frac{d^2\epsilon(q)}{dq^2}(G_K-G_R+G_A)(q,q_0).
\end{align}
Here we have modified the expression of $D_d$ by adding the $G_R-G_A$ because of the subtlety in Keldysh approach \cite{Kamenev} when we use the Green's function $G(t,t)$ with two time arguments coincident. This is indeed correct because we know for $\epsilon(k)=k^2$ the result should be proportional to charge density while by fluctuation-dissipation theorem, $G_K$ is proportional to $(1-2n_F)$ and we need to add a term to cancel the constant "$1$" here.

The first line of \eqref{trans S} can be written as $\mathcal{J}^\mu\partial_\mu\phi$ where $\mathcal{J}^\mu$ is the $U(1)$ current for charge in free-fermion limit. The contribution to the action is calculated by evaluating the bubble in Figure \ref{trans} which is a bubble for  current-current correlation function:

\begin{align}
iS^{(3)}_{\mathcal{J}}=&(D_{pp})\partial_x\phi_q\partial_x\phi_{\text{cl}}+D_{\omega\omega}(\partial_t\phi_{\text{cl}}+J_{\text{cl}})(\partial_t\phi_{q}+J_{q})\notag\\+&D_{\omega p}\left(\partial_x\phi_{\text{cl}}(\partial_t\phi_{q}+J_{q})+(\partial_t\phi_{\text{cl}}+J_{\text{cl}})\partial_x\phi_{q}\right).
\end{align}
with:
\begin{align}
D_{\zeta_1\zeta_2}(p,\omega)=&\notag\\-\int\frac{dq_0dq}{(2\pi)^2}(&G_R(q+\frac{p}{2},q_0+\frac{\omega}{2})G_K(q-\frac{p}{2},q_0-\frac{\omega}{2})\notag\\&G_K(q+\frac{p}{2},q_0+\frac{\omega}{2})G_A(q-\frac{p}{2},q_0-\frac{\omega}{2}))\notag\\&\left(\frac{d\epsilon(q)}{dq}\right)^{\delta_{\zeta_1,p}+\delta_{\zeta_2,p}}.
\end{align}
Where we have used a simplified notion. Summing up all contributions, the action for the retarded Green's function is given by:
\begin{align}
iS_{q,\text{cl}}[\phi,J]=&(D_r+D_d+D_{pp})\partial_x\phi_q\partial_x\phi_{\text{cl}}\notag\\+&D_{\omega p}\left(\partial_x\phi_{\text{cl}}(\partial_t\phi_{q}+J_{q})+(\partial_t\phi_{\text{cl}}+J_{\text{cl}})\partial_x\phi_{q}\right)\notag\\+&D_{\omega\omega}(\partial_t\phi_{\text{cl}}+J_{\text{cl}})(\partial_t\phi_{q}+J_{q}).
\end{align}
as a result, 
\begin{align}
\Pi_{R,nn}&=-\frac{i}{2}\left(D_{\omega\omega}+\frac{-(\omega D_{\omega\omega}-pD_{\omega p})^2}{D'_{pp}p^2+D_{\omega\omega}\omega^2-2\omega p D_{\omega p}}\right). \label{Pinn}
\end{align}
Where $D$s here may depend on $\omega$ and $p$ and $D'_{pp}=D_r+D_d+D_{pp}$. 

We first would like to check this expression \eqref{Pinn} reproduce the correct formula in known limits at least small $p$ and $\omega$. With $V=J=0$, one could use the non-interacting result for $G_R$ and one finds $$\Pi_{R,nn}=-\frac{i}{2}D_{\omega\omega}(p,\omega)=\Pi_{R,nn}^{DM}(p,\omega).$$ thanks to the fact $$\omega D_{\omega\omega}-pD_{\omega p}=0,$$ for $\omega$, $p\rightarrow0$. In fact, for free fermions, the only contribution comes from the wick contraction of two density operators and the Wald identity $\omega D_{\omega\omega}-pD_{\omega p}=0$ should be always true even for large $p$ and $\omega$ if we do't approximate the current operator by its low-energy form. This fact, in language of the standard effective action \cite{Comments,thermal transport,Balent} for (generalized) SYK models, means that the phase fluctuation we considered here should be a fluctuation of self energy $\Sigma(t-t')\rightarrow\exp(-i\phi(t))\Sigma(t-t')\exp(i\phi(t))$. But for $\Sigma=0$, there is no such mode and the density $n$ should not couple to $\phi$ because now $\phi$ is only a unphysical gauge transformation. Indeed mathematically this argument leads to the Wald identity for charge conservation.

In the opposite limit, if $t=0$, then we should set $D_{\omega p}=D_{p p}=D_d=0$ and we have:
\begin{align}
\Pi_{R,nn}=-\frac{i}{2}\left(\frac{D_{\omega\omega}D_r p^2}{(D_r)p^2+D_{\omega\omega}\omega^2}\right)=\Pi_{R,nn}^{HD}. \label{PinnHD}
\end{align}
if one approximate: $$D_{\omega\omega}(p,\omega)\sim2i K,\ \ D_{r}(\omega)\sim-2\omega K D,$$ as in \cite{Balent}. 

\begin{figure}[t]
	\centering
	\includegraphics[width=0.45\textwidth]{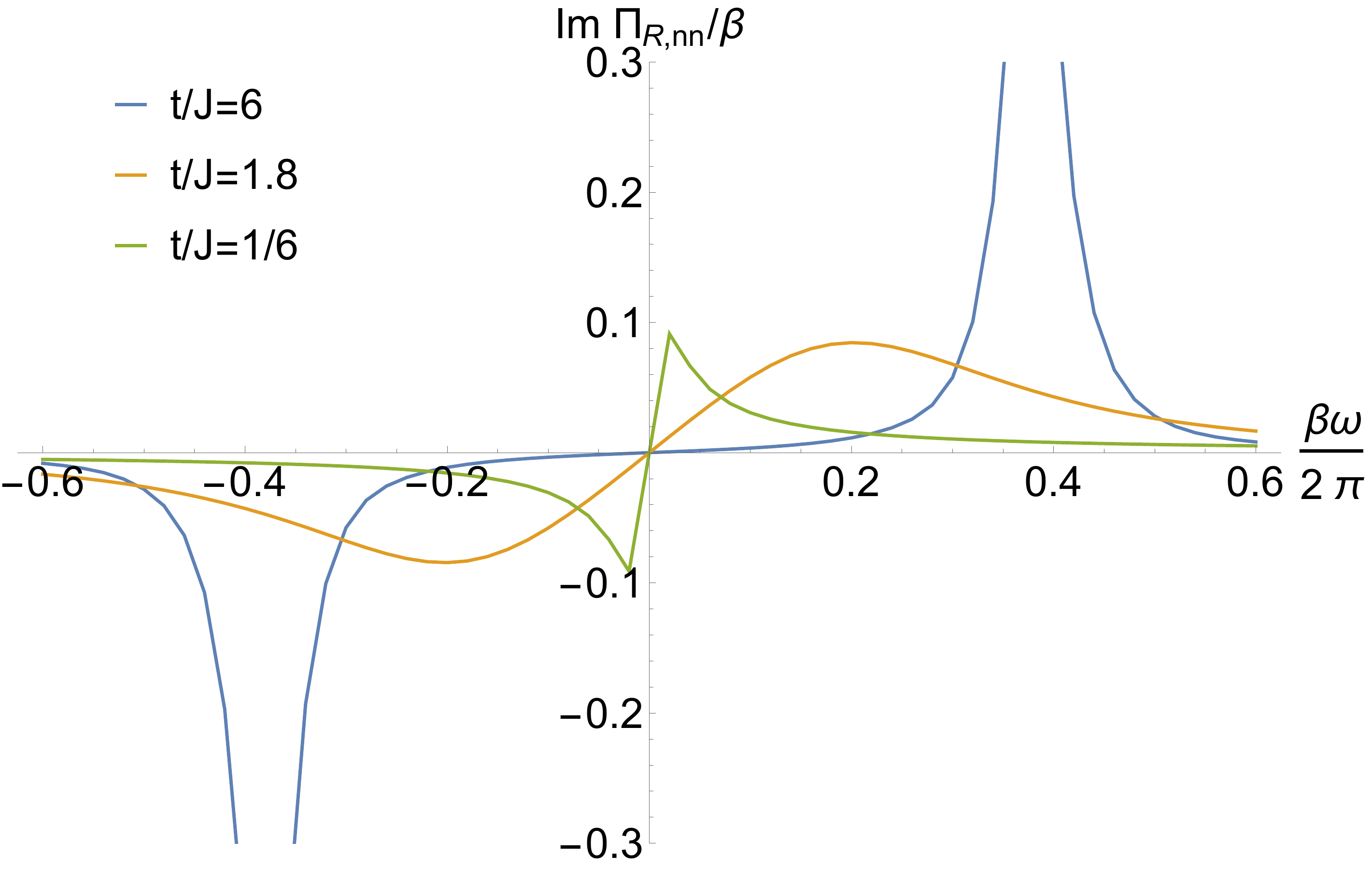}
	\caption{The spectral for density correlation function $\text{Im}\Pi_{R,nn}(\omega,0.1)$ for different $J$ and $t$ with $V=J$, $\mu=0$ and $\beta t+\beta J=12$.}\label{Pinnfig}
\end{figure}

Now we would like to present the numerical result for the crossover between these two limits. Using the numerical results for two-point correlation function, we calculate the $\Pi_{R,nn}$. We plot the imaginary part $\text{Im}\Pi_{R,nn}(\omega,0.1)$ in Figure \ref{Pinnfig} for $\mu=0$ and $V=J$. For small $J$ and large $t$, we see two peaks at $\pm2t p$ which is the density mode. Tuning $t$ to be smaller and $J$ larger, the two peak get closer and becomes much broader. Eventually they touch the zero point and becomes a single dissipative mode. As we said before, this may be a starting point for the construction of a bulk dual, which now is still unclear. Because we know the dispersive metal phase is not maximally chaotic, we know this crossover should be embedded in a string theory whose strongly(weakly) interacting limit gives the dispersive metal (incoherent metal) phase.

\section{Lyapunov exponent and band structure}
It is well known that the non-Fermi Liquid phase described by SYK model is maximally chaotic\cite{Comments,Kitaev2}. Then it is natural to study the effect of dispersion on quantum chaos using our dispersive SYK model. The chaotic behavior is characterized by Lyapunov exponent $\lambda_L$  which is defined in the introduction. The calculation of $F(t)$ can be done by first calculating the four-point correlation function in the imaginary time path-integral and then do the analytical continuation. But here we the choose more direct way by using the Keldysh approach with four time contour \cite{Comments}.

For dispersive SYK model, because of the U(1) symmetry, we have to define two different types of correlator which couple together as in \cite{Altman}. Moreover, because the fermions can now hopping, so the OTOC of different sites will all couple directly, compared to only directly couplings between the nearest neighbor sites in randomly coupled SYK models \cite{Yingfei1,Yingfei2,thermal transport}.

\begin{figure}[t]
	\centering
	\includegraphics[width=0.45\textwidth]{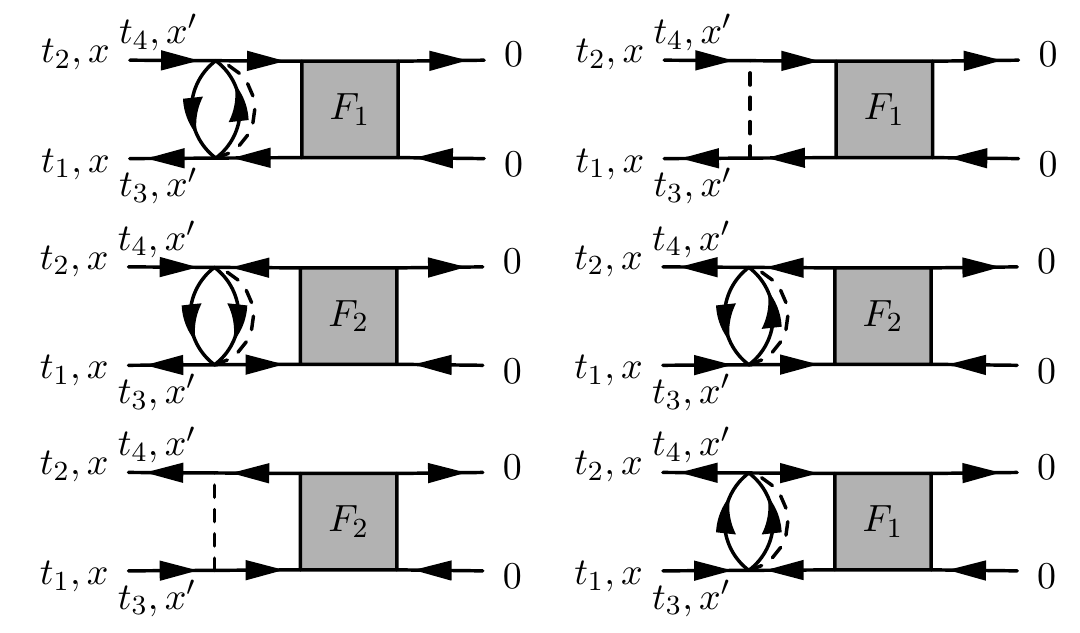}
	\caption{Diagrammatic representation of the self-consistent equation for OTOC. The OTOC at $x$ will couples to OTOC at any $x'$ directly because there is spatial correlation for two-point Green's function.}\label{consistent}
\end{figure} 

We define:
\begin{align}
F_1(t_1,t_2,x)=\left<c_x^\dagger(t_2-i\frac{3\beta}{4})c_0(0-i\frac{\beta}{2})c_x(t_1-i\frac{\beta}{4})c_0^\dagger(0)\right>_\beta,\\
F_2(t_1,t_2,x)=\left<c_x(t_2-i\frac{3\beta}{4})c_0(0-i\frac{\beta}{2})c_x^\dagger(t_1-i\frac{\beta}{4})c_0^\dagger(0)\right>_\beta.
\end{align}

As shown in the Figure \ref{consistent}, the self consistent equation for $F_1$ and $F_2$ is given by:
\begin{align}
F_1&(t_1,t_2,x)=\notag\\&\int dt_3dt_4\sum_{x'}(K_{11}+K_{13})(t_1,t_2,x;t_3,t_4,x')F_1(t_3,t_4,x')\notag\\&\ \ \ \  +K_{12}(t_1,t_2,x;t_3,t_4,x')F_2(t_3,t_4,x'),\\
F_2&(t_1,t_2,x)=\notag\\&\int dt_3dt_4\sum_{x'}(K_{22}+K_{24})(t_1,t_2,x;t_3,t_4,x')F_2(t_3,t_4,x')\notag\\&\ \ \ \ +K_{21}(t_1,t_2,x;t_3,t_4,x')F_1(t_3,t_4,x').
\end{align}
with $K_{ab}(t_1,t_2,x;t_3,t_4,0)$ given by:
\begin{align}
&K_{11}=-2J^2G^*_R(x,t_{24})G_R(x,t_{13})G_{lr}^+ (t_{34})G_{lr}^+ (t_{43}),\\
&K_{13}=\frac{V^2}{2}\sum_\pm G^*_R(x\pm1,t_{24})G_R(x\pm1,t_{13}),\\
&K_{12}=-J^2G^*_R(x,t_{24})G_R(x,t_{13})G_{lr}^+ (t_{34})G_{lr}^+ (t_{34}).
\end{align} 
and 
\begin{align}
&K_{22}=-2J^2G_R(x,t_{24})G^*_R(x,t_{13})G_{lr}^+ (t_{34})G_{lr}^+ (t_{43}),\\
&K_{24}=\frac{V^2}{2}\sum_\pm G_R(x\pm1,t_{24})G^*_R(x\pm1,t_{13}),\\
&K_{12}=-J^2G_R(x,t_{24})G^*_R(x,t_{13})G_{lr}^+ (t_{43})G_{lr}^+ (t_{43}).
\end{align}
Here we take the long time limit and keep only homogeneous terms. We have defined $G_{lr}^+ (t)=-i\left<c_x(t-i\frac{\beta}{2})c_x^\dagger(0)\right>_\beta$ and used $G_R(t,x)=G_A^*(-t,-x)$. The kernel of the self-consistent equation is invariant under time translation $t_i\rightarrow t_i+a$ where $a$ is a constant and at long time limit, we expect for a chaotic system the time translation generator has a imaginary eigenvalue \cite{Comments}. We assume:
\begin{align}
F_i(t_1,t_2)=\exp(\lambda_L(t_1+t_2)/2)f_i(t_1-t_2).
\end{align}
\begin{figure}[t]
	\centering
	\includegraphics[width=0.45\textwidth]{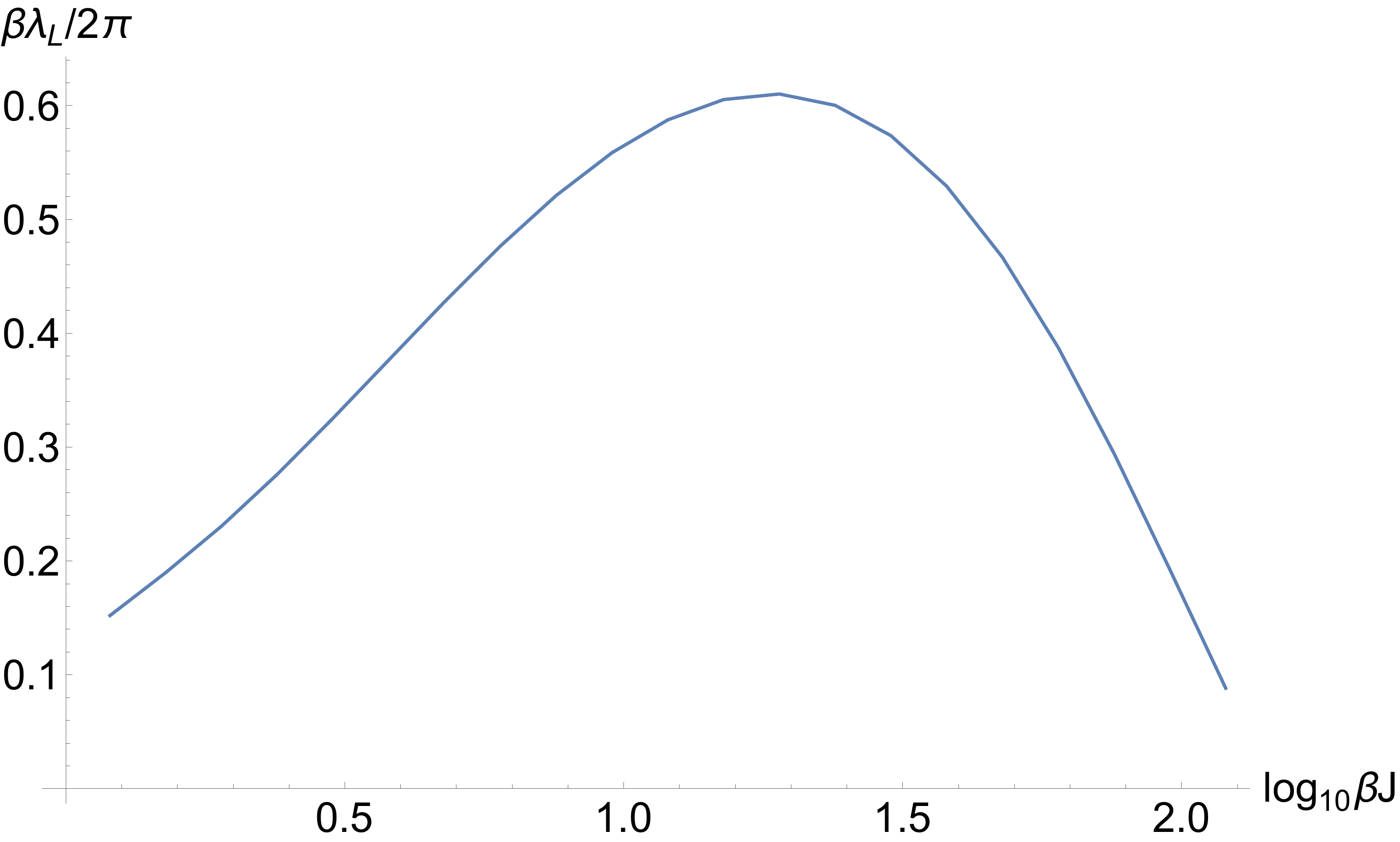}
	\caption{Lyapunov exponent for different temperature. We fix $J/t=10$, $\mu=V=0$ and only consider $p=0$. The Lyapunov exponent shows a crossover between a low-temperature weakly interacting phase and a higher temperature SYK phase.}\label{lyatem}
\end{figure} 

Because of the time and spatial translational invariance, we'd better go to frequency and momentum domain. As a result, the equation is given by:
\begin{align}
f_1&(p,\omega)=G^s_R(p,\omega)(V^2\cos(p)f_1(p,\omega)\notag\\&+J^2\int d\omega'\left(2g^1_{lr}(\omega-\omega')f_1(\omega')+g^2_{lr}(\omega-\omega')f_2(\omega')\right)),
\\
f_2&(p,\omega)=G^s_R(p,-\omega)(V^2\cos(p)f_2(p,\omega)\notag\\&+J^2\int d\omega'\left(2g^1_{lr}(\omega-\omega')f_2(\omega')+g^3_{lr}(\omega-\omega')f_1(\omega')\right)).
\end{align}
where we defined:
\begin{align}
&G^s_R(p,\omega)\equiv\sum_x\exp(ipx)|G_R(x,\omega+\frac{i\lambda_L}{2})|^2,\\
&g^1_{lr}(\omega)=-\int\frac{dt}{2\pi}G^+_{lr}(t)G^+_{lr}(-t)\exp(i\omega t),\\
&g^2_{lr}(\omega)=-\int\frac{dt}{2\pi}G^+_{lr}(t)G^+_{lr}(t)\exp(i\omega t),\\
&g^3_{lr}(\omega)=-\int\frac{dt}{2\pi}G^+_{lr}(-t)G^+_{lr}(-t)\exp(i\omega t).
\end{align}

One could use the relation $G^+_{lr}(\omega)=\frac{-iA(x=0,\omega)}{2\cosh(\beta\omega/2)}$ to calculate $G_{lr}^+$. To calculate $G_R(p,\omega+i\lambda_L/2)$, one could go back to \eqref{SDeGR} and directly derive the self-consistent equation with shifted frequency:
\begin{align}
G^{-1}_R(p,\omega+\frac{i\lambda_L}{2})&=\omega+\frac{i\lambda_L}{2}-\epsilon(p)-V^2G_R(\omega+\frac{i\lambda_L}{2})\notag\\&-J^2\tilde{\Sigma}_R(\omega),
\end{align}
where we have: 
\begin{align}
\tilde{\Sigma}_R(\omega)&=-iJ^2\int_0^\infty dt e^{i\omega t-\frac{\lambda_L t}{2}}(n_1^2(t)n_2(t)+n_3^2(t)n_4(t)).
\end{align} 

We first show the crossover between different fixed points when tuning temperature. As discussed before, the low-temperature physics is captured by fermions hopping on a lattice with random interaction, which has quasi-particle and should be slow-scrambling. If $t\ll J$, then if we increase the temperature, we will get into the regime dominated by SYK fixed point and this will be a fast-scrambling non-Fermi Liquid without quasi-particle. As shown in Figure \ref{lyatem}, we find the expected non-monotonic behavior for $\lambda_L$ with $p=0$. In high temperature limit, this curve coincident with the result for the original SYK model \cite{Comments}.

\begin{figure}[t]
	\centering
	\includegraphics[width=0.45\textwidth]{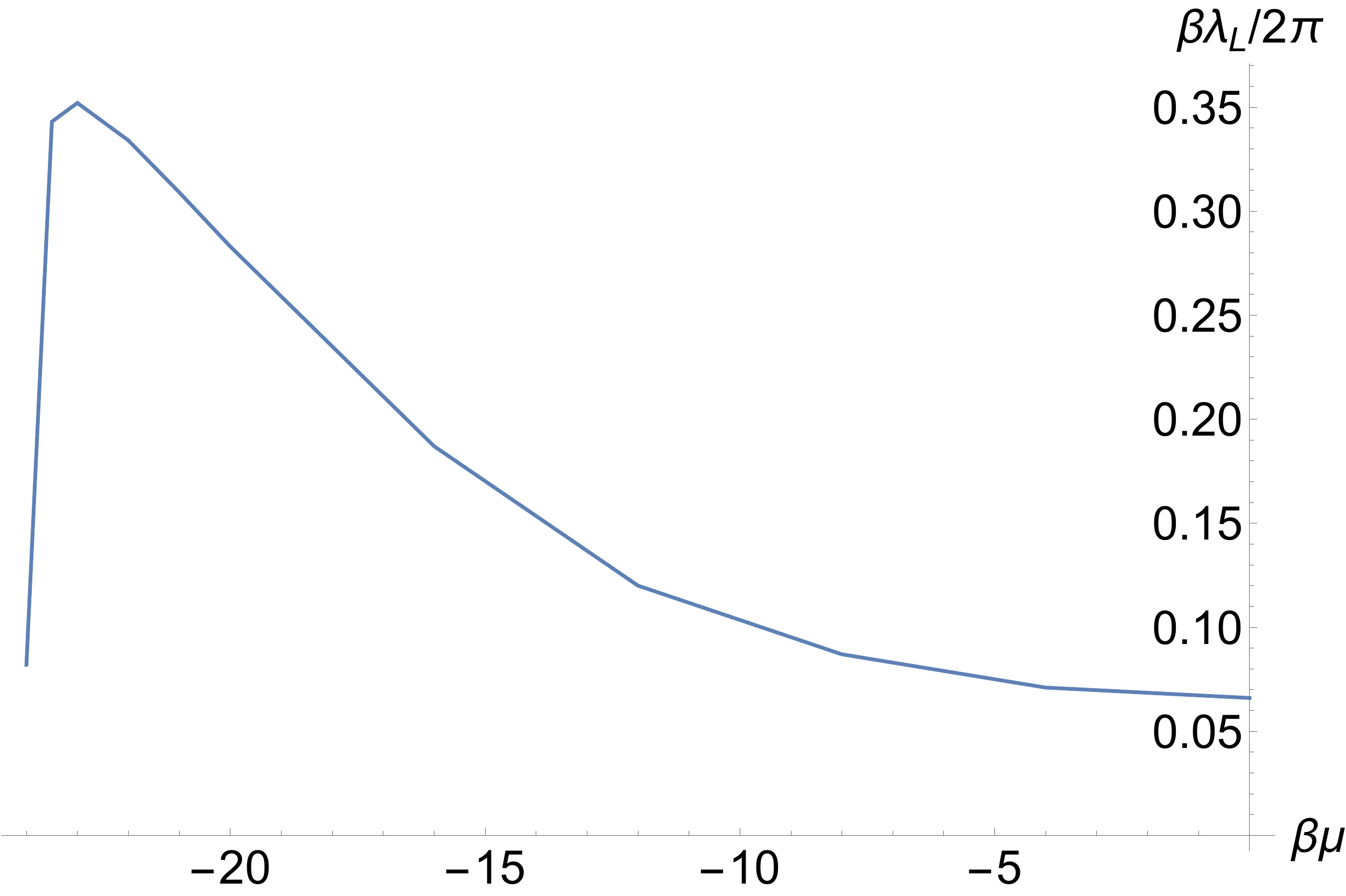}
	\caption{Lyapunov exponent for fillings. We fix $\beta J=50$, $\beta t=10$ and $V=0$. We only consider $p=0$. The chaotic behavior largely enhanced when $\mu$ approach the Van Hove singularity.}\label{lyamu}
\end{figure}
We also study the effect of band-structure on Lyapunov exponent. As explained in section 2, we indeed find that near a Van Hove singularity, $\lambda_L$ is largely enhanced because more degrees of freedom are included in the long-time limit. In this sense, traditional SYK model, which contains the largest possible density of states at $\omega\sim0$ without the chaotic random interaction (numbers of delta peaks), should be maximally chaotic.

One should then expect another possibility of enhancing quantum chaos by the topology of the single-particle band. For example, first set $V=\mu=0$. If one have a SSH model with hopping strength $t$ and $t'$, the band is topologically non-trivial if $t<t'$. Then for a open chain there will be a localized edge state, and the local density of states is very large. We then expect the Lyapunov exponent to get a spatial dependence which is largest near the edge. For the special case $t=0$, the edge mode then locals at a single site and on this site the model reduces exactly to a SYK quantum dot with maximally chaotic behavior.

	\section{Summary and Outlook}
	In this paper, we studied the dispersive SYK model by calculating spectral function, density correlation function and Lyapunov exponent. By either tuning the relative strength of constant hopping and random interaction or temperature, we find a crossover between a dispersive metal and an incoherent metal without dispersion. We also find the Lyapunov exponent is largely affected by the density of states near the Fermi surface.
	
	We would like to discuss some possible extension of this work. In \cite{Altman,sk jian,our}, the authors consider coupling SYK sites with different number of modes and this leads to a new non-Fermi Liquid fixed point which is also maximally chaotic and stable at low temperature. It will be interesting to generalize dispersive SYK model to cover the case with different number of modes on each site, which may lead to new crossover physics.
	
	Effect of an additional constant interaction may also be interesting. For example, considering a time-reversal invariant system with attractive interaction together and a band-structure, there is a cooper instability. As a result, in the low temperature limit, for high dimension, the fermions should pair and forms a superconductor. At higher temperature, the band structure is washed out by the random interaction and this induce a transition between superconductor and incoherent metal. The critical behavior of this transition may be interesting.
	
	Dispersive SYK model also provide a platform to study the interplay between topology of the band-structure and quantum chaos in more details than the brief discussion in the last section. For example, in $2+1$-$d$ one may study coupling Haldane model lattice by random interaction, which may give an example of topologically non-trivial insulator with strong interaction. The possibility of some fractional Chern insulator may also be interesting. We defer these for further study.
	
	\textit{Acknowledgment.} We thank Hui Zhai, Chao-Ming Jian, Yu Chen, Ruihua Fan, Yiming Chen and Xin Chen for helpful discussion.
	

\end{document}